\documentclass{elsart}

\usepackage{graphicx,epsfig}

\makeatletter

\def\elsartstyle{%

        \def\normalsize{\@setfontsize\normalsize\@xiipt{14.5}}

        \def\small{\@setfontsize\small\@xipt{13.6}}

        \let\footnotesize=\small

        \def\large{\@setfontsize\large\@xivpt{18}}

        \def\Large{\@setfontsize\Large\@xviipt{22}}

        \skip\@mpfootins = 18\p@ \@plus 2\p@

        \normalsize

}

\makeatother

\def\url#1{{\ttfamily\def\/{/\discretionary{}{}{}}#1}}

\begin{document}
\begin{center}
talk given at the School \\
{\it Understanding our Universe at the close of 20th Century}
\\
25 April 2000 - 6 May 2000
\\
Cargese, France
\end{center}
\vskip2mm
\hrule
\begin{frontmatter}

\title{Supernovae and Cosmology}
\author{Monique Signore$^\star$, Denis Puy$^{\dagger,\ddagger}$}
\address{$^\star$Laboratoire de Radioastronomie, Observatoire de Paris 
(France)
\\
$^\dagger$Paul Scherrer Institut, Villigen (Switzerland)
\\
$^\ddagger$Institute of Theoretical Physik, University of Zurich (Switzerland)}
\thanks[email]{E-mail: monique.signore@obspm.fr, puy@physik.unizh.ch}
\begin{abstract}
These lecture notes intend to form a short pedagogical introduction to the 
use of {\it typical} type Ia-Supernovae (hereafter SNIa) as standard candles 
to determine the energy density of the universe. Problems of principle 
for taking SNIa as cosmological probes are pointed out, and new attempts 
at solving them are indicated including the empirical width-luminosity 
relation (WLR) and its possible explanations.
\\
Finally, the observations of SNIa at high redshift carried out by two major 
teams are briefly reviewed and their interpretation as evidence for an 
accelerating universe is also rapidly discussed 
\end{abstract}
\end{frontmatter}
\section{Introduction}
A supernova is a very powerful event: a star which suddenly brightens to about 
10$^9$-10$^{10}$ L$_\odot$. There are two different types of supernovae 
(see Table 1): SNII and SNIa, such as:
\begin{itemize}
\item SNII are core-collapse induced explosions of short-lived massive stars 
\\
(M$_\star \, > $ 8 M$_\odot$) which produce more $O$ and $Mg$ relative to $Fe$.
\item SNIa are thermonuclear explosions of accreting white-dwarfs in close 
binaries which produce mostly $Fe$ and little $O$; the exact companions 
stars of white-dwarfs are generally not identified but must be relatively long-lived stars.
\end{itemize}
There are also particular supernovae: $SNIb$ and $SNIa$ with most properties 
of $SNII$ except their spectra which have no $H$ line (such as SNIa) but they 
believed to be core-collapse supernovae (such as SNII).
\\

\begin{table*}[h!tbp]
 \renewcommand{\arraystretch}{.90}
  \centering
    \begin{tabular}{||l||c|c||}\hline \hline
 {\bf Type} & {\bf SNIa}  & {\bf SNII}  \\
\hline \hline 
 & & \\
{\bf Hydrogen} & no & yes \\
\hline 
 & & \\ 
{\bf Optical spectrum} & Metal lines & P Cyg lines \\
 & deep 6180 $\AA$ & Balmer series \\
\hline 
 & & \\
{\bf Absolute luminosity} & $\sim$ 4$\times 10^9$ 
L$_{\odot}$ & $\sim$ 10$^9$ L$_\odot$ \\
at max. light & small dispersion  & large dispersion \\
 & standard candles ? & \\ 
\hline  
 & & \\
{\bf Optical light curve} & homogeneous & heteregeneous 
\\
\hline  
 & & \\
{\bf UV spectrum} & very weak & strong \\
\hline 
  & & \\
{\bf Radio emission} & no detection & strong \\ 
 & & slow decay\\
\hline  
 & & \\
{\bf Location} & all galaxies & spirals and irregulars\\
\hline 
  & & \\
{\bf Stellar population} & old & young\\
\hline  
 & & \\
{\bf Progenitors} & WD in binary systems & massive stars\\ 
\hline \hline
    \end{tabular}
    \label{tab:geom}
\end{table*}

For some time astronomers had focused on all types of supernovae to measure Hubble's constant. Now, with the use of Hubble Space Telescope (HST) 
observations of Cepheids in the host galaxies of these supernovae, there is 
a consensus that $H_o$ is in the range of 60-70 km s$^{-1}$ Mpc$^{-1}$ -see 
Branch (1998) for a review.
\\
But very recently, two observational groups working independently to use SNIa as 
standard candles to measure distance versus redshift presented evidence that 
the Hubble expansion has been accelerating over time. In effect, these two 
independent groups (Perlmutter et al. 1998, Schmidt et al. 1998) had to 
develop not only a method of discovering SNIa at high redshift by careful 
observations -with large format CCDs, large apertures- and scanning of plates 
but they also had to determine if these SNIa could be used as standard candles. This recent and important progress has been possible because an empirical 
relation (Phillips 1993) between the duration of the peak phase of a 
SNIa's light curve and its luminosity ({\it broader is brighter}) has been 
taken into account by both teams.
\\
This is precisely the work of these two groups that we are trying to present, in an educational manner, in these lessons. In this scope, in section 2, we 
summarize our current understanding of SNIas: their possible progenitors; some 
models of their explosions; observed optical spectra of SNIas, $\gamma$-rays 
from SNIas, observed optical light curves and finally the empirical width-
luminosity relation (WLR) which is also called the {\it Phillips Relation}. 
In section 3, we give a cosmological background by recalling what is the 
($\Omega_\Lambda$, $\Omega_M$)-plane and the luminosity distance. In section 4, we present the 
observations of SNIa at high redshift by the two major teams: the 
{\it Supernovae Cosmology Project} (SCP) and the {\it High z supernova search 
team} (HZT). In section 5, we discuss some aspects of their results and 
finally give a conclusion.
\section{On type Ia supernovae}
There are spectroscopic and photometric indications that SNIa result from the 
thermonuclear explosions of accreting carbon/oxygen white dwarfs (hereafter 
WD). However, the progenitor systems of SNIa -their nature, their evolution- 
the hydrodynamical models for SNIa- the mass of the WD at ignition, the physics
 of the nuclear burnings- are still uncertain.
\\
Moreover, because they are among the brightest optical explosive events in 
the universe, and because their light curves and their spectral evolution 
are {\it relatively uniform}, SNIa have been tentatively used as standard 
candles. However variations of light curves and spectra among SNIa have 
recently been extensively studied; in particular, the relation between the 
duration of the peak phase of their light curves and their luminosities -
{\it broader and brighter}- Phillips (1993).
\\
This section briefly summarizes our knowledge of these SNIa properties and 
also their controversial issues. However, for more insight into the underlying
 physics of SNIa and their progenitors, see the various excellent reviews 
of Woosley-Weaver (1986), Nomoto et al. (1997), Livio (2000) and most of the 
papers of the following proceedings {\it Supernovae} (Petscheck, 1990), 
{\it Thermonuclear Supernovae} (Ruiz-Lapente et al. 1997); 
{\it Type Ia Supernovae and Cosmology} (Niemeyer Truran 2000), {\it Cosmic 
Explosions} (Holt \& Zhang 2000). 
\subsection{On the progenitors of SNIa}
There are two classes of models proposed as progenitors of SNIa:
\vskip2mm
{\it i)} The Chandrasekhar mass model in which a mass acreting $C+O$ WD grows in mass up to the critical mass: $M_{Ia} \sim 1.37-1.38$ M$_\odot$ -which is 
near the Chandrasekhar mass and explodes as a SNIa.
\\
{\it ii)} The sub-Chandrasekhar mass model, in which an accreted layer of 
helium, atop a $C+O$ WD ignites off-center for a WD mass well below the 
Chandrasekhar mass. 
\\
The early time spectra of the majority of SNIa are in excellent agreement 
with the synthetic spectra of the Chandrasekhar mass models. The spectra of 
the sub-Chandrasekhar mass models are too blue to be comparable with the 
observations. But let us only give some crude features about the progenitor 
evolution.
\subsubsection{Chandrasekhar mass models}
For the evolution of accreting WD toward the Chandrasekhar mass model, two 
scenarii have been proposed: a) a double degenerate (DD) scenario, i.e. the 
merging of the $C+O$ WD with a combined mass surpassing the Chandrasekhar mass 
$M_{ch}$; b) a single degenerate (SD) scenario, i.e. the accretion of 
hydrogen rich matter via mass transfer from a binary companion. The issue of 
DD versus SD scenario is still debated. Moreover, some theoretical modeling 
has indicated that the merging of WD lead to the accretion-induced collapse 
rather than a SNIa explosion.
\subsubsection{Sub-Chandrasekhar mass models}
In the sub-Chandrasekhar mass model for SNIa, a WD explodes as a SNIa 
\underline{only} when the rate of the mass accretion rate $\dot{M}$ is in 
a certain narrow range. Moreover, for these models, if $\dot{M}>\dot{M}_{crit}$,
 a critical rate, the accreted matter extends to form a common envelope. 
However, this difficulty has been recently overcome by a WD wind model. For 
the present binary systems which grow the WD mass to $M_{Ia}$ there are two 
possible systems:
\vskip2mm
{\it a)} a mass-accreting WD and a lobe-filling main sequences star (WD+MS system).
\\
{\it a)} a mass-accreting WD and a lobe-filling less massive red giant star 
(WD+RG system).
\vskip2mm
Let us conclude this subsection by saying that the \underline{evolution of the 
progenitor} is undoubtly the most uncertain part of a fully predictive model 
of a SNIa explosion.
\subsection{On explosion models}
As already noted above, the first important question is about {\it 
ignition}, which can be formulated as: {\it when and how does the burning 
start ?} We have also seen that there are eventually two candidates:
\vskip2mm
{\it i)} the ignition of $^{12}C+^{12}C$, in the core of a WD composed of 
$C+O$, is due to compressive heating from accretion; this is the case of 
Chandrasekhar models, for which: $M_{WD} \sim M_{ch} \sim 1.4 $ M$_\odot$.
\\
{\it ii)} the ignition of $He$ in a shell around a $C+O$ core is followed by 
the explosive burning of the $C+O$ core and the helium shell; this is the case 
of sub-Chandrasekhar mass models for which 0.7 M$_\odot < M_{WD} < 1.4$ 
M$_\odot$.
\vskip2mm
The second important question is {\it once ignited, does the flame propagate 
supersonically by detonation (new fuel heated by shock compression ) ? or 
subsonically by deflagration (new fuel heated by conduction) ?}.
\\
The physics involved in these processes is very complex and 
relative to the physics of 
thermonuclear combustion. In particular, the physics of turbulent combustion 
and the possible spontaneous transition to detonation are probably the most 
important and least tractable effects. Multidimensional simulations are 
presently conducted to understand these processes. 
\\
Because this work is a pedagogical introduction to the use of SNIa as cosmological probes, let us \underline{only} consider Chandrasekhar mass models as 
{\it standard SNIa} models: they provide a point of convergent evolution for 
various progenitor systems and therefore could offer a natural explanation of 
the assumed uniformity of display. Here, we only mention two Chandrasekhar 
mass models (groups of Nomoto and Woosley) that can account for the basic 
features of the so-called {\it standard SNIa}, i.e. : i) an explosion energy 
of about 10$^{51}$ ergs; ii) the synthesis of a large amount of $^{56}Ni$ 
($\sim 0.6$ M$_\odot$); iii) the production of substantial amounts of intermediate-mass elements at expansion velocities of about 10 000 km s$^{-1}$ near the maximum brightness of SNIa explosions. All these features are very important 
to explain observed optical displays of SNIa: spectra and light curves.
\subsubsection{The carbon deflagration model W7}
This model (Nomoto et al. 1984) has been seen for a long time as the standard 
explosion model of SNIa. In this model, a subsonic deflagration wave 
propagates at an average speed of 1/5 of the sound speed from the center to the expanding outer layers. The explosion synthesis of $\sim 0.58$ M$_\odot$ 
of $^{56}Ni$ in the inner region is ejected with a velocity of 1-2 
$\times$ 10$^4$ km s$^{-1}$.
\subsubsection{Delayed detonation models}
Since 1984, there has been great progress in applying concepts from terrestrial
 combustion physics to the SNIa problem. In particular, it has been shown that the outcome of carbon deflagration depends critically on its flame velocity 
which is still uncertain. If the flame velocity is much smaller than in W7, 
the WD might undergo a pulsating detonation; the deflagration might also 
induce a 
detonation, for instance model DD4 of Woosley \& Weaver (1994) or model 
WDD2 of Nomoto et al. (1996).
\\
Thus the full details of the combustion are quite uncertain; in 
particular: progenitor evolution, ignition densities, effective propagation, 
speed of the burning front, type of the burning front (deflagration, 
detonation, both at different stages and their transitions). 
\\
However, constraints on these still uncertain parameters in models such as 
W7, DD4, WDD2 can be provided by comparisons of synthetical spectra and 
light curves with observations, by comparisons of predictive nucleosynthesis 
with solar isotopic ratio.
\subsection{On the observations of SNIa}
Here, we only review the observations of SNIa which are relevant for our 
cosmological problem which can be formulated by the following question: 
Can one consider SNIa as a well-defined class of supernovae in order to 
use them as {\it standard candles ?} Therefore, we briefly present some of the 
observed optical spectra and optical light curves of the so-called SNIa. 
Moreover, we also give some considerations on $\gamma$-rays from SNIa. 
\subsubsection{On optical spectra of SNIa}
Optical spectra of SNIa are generally homogeneous but one can also observe 
some important variations. One recalls that SNIa have not a thick $H$-rich 
envelope so that the elements which are synthesized during their explosions 
must be observed in their spectra. Therefore, \`a-priori, a comparison between 
synthetic spectra and observations can be a powerful diagnostic of dynamics 
and nucleosynthesis of SNIa models. Spectra have been calculated for various 
models and can be used for comparisons with observations. For instance: 
\vskip2mm
{\it i)} Nugent et al. (1997) present agreement between observed spectra of SN 1992A (23 
days after the explosion), SN 1994A (20 days after the explosion) and 
synthetic spectra of the W7 model for the same epochs after the explosions.
\\
Moreover, the various spectral features can be identified as those of $Fe$, 
$Ca$, $S$, $Si$, $Mg$ and $O$. As already noted above, synthetic spectra for 
the sub-Chandrasekhar mass models are less satisfatory.
\\
{\it ii)} For heterogeneity Nomoto et al. (1997) present observed 
optical spectra of SNIa (SN 1994D, SN 1990N) about one week before maximum 
brightness which show different features from those of SN1991T observed at 
the same epoch. Let us note also that SN1991T is a spectroscopically peculiar 
SNIa. 
\subsubsection{On $\gamma$-rays from SNIa}
SNIa synthesise also radioactive nuclei such as $^{56}Ni$, $^{57}Ni$, 
$^{44}Ti$. The species $^{56}Ni$ is certainly the most abundant radioactivity 
produced in the explosion of a SNIa. As we will wee below, the amount of 
$^{56}Ni$ ($\sim 0.6$ M$_\odot$) is very important for the physics of the 
optical light curve of the SNIa. But in fact, one must consider the following 
decay chain:
$$
^{56}Ni \, 
\stackrel{{\rm 9 \ days }}{\longrightarrow} \, ^{56}Co \,  
\stackrel{{\rm 112 \ days}}{\longrightarrow} \, ^{56}Fe + e^+
$$
And, for a certain time after the explosion the hard electromagnetic spectrum 
is dominated by $\gamma$-rays from the decays of $^{56}Co$. During each 
decay of $^{56}Co$ a number of $\gamma$-rays lines are produced. The average 
energy is about 3.59 MeV including 1.02 MeV from $e^+-e^-$ annihilations. 
The most prominent lines are the 847 keV line and the 1238 keV line. After 
$^{56}Co$ decays, other unstable nuclei become also important both for the 
energy budget and for producing observable hard emissions. One must calculate,
 for every SNIa model, the $\gamma$-ray light curve, the date, the flux at 
the maximum of the light curve for the 847 keV line and 1238 keV line -see for 
instance many {\it $\gamma$-rays papers} in Ruiz-Lapente et al. 1997).
\vskip2mm
From the observational point of view:
\begin{itemize}
\item i) SN 1986G (in NGC 5128) was observed by SMM (Solar Maximum Mission). 
An upper limit of: $F_{847} \leq 2.2 \times 10^{-4} \gamma$ cm$^{-2}$ s$^{-1}$ 
was reported leading to an upper limit of 
$$
M_{56}(3\sigma) < 0.4 \Bigl( \frac{D}{3{\rm \, Mpc}} \Bigr)^2 \, {\rm M}_\odot.
$$
Later, SN 1986 G was estimated as an under luminous SNIa (Phillips 1993).
\item ii) SN1991T (in NGC 4527) was observed by CGRO (Compton Gamma Ray 
Observatory) on day 66 after the explosion. A detection by the instrument COMPTEL/CGRO of the 847 keV line has been reported with a flux of $F_{847}=5.3 
\pm 2.1 \times 10^{-5} \gamma$ cm$^{-2}$ s$^{-1}$. During the same date, 
no detection, by the instrument OSSE/CGRO, of the 847 keV line has been 
reported. A 3$\sigma$-upper limit  $F \leq 4.5 \times 10^{-5} \gamma$ cm$^{-2}$
 s$^{-1}$ has been given only (see the {\it $\gamma$-ray papers} in Ruiz-Lapente
 et al. 1997 and references therein).
\\
Let us also recall that SN 1991 T was quoted as a peculiar SNIa, from the 
point of view of optical spectroscopy (see above 2.3.1).
\item iii) An interesting test for the nature of SNIa explosion models has 
been found and published just after Cargese 2000, Pinto et al. (2000) present 
results of $X$ and $\gamma$-ray transport calculations of M$_{ch}$ and 
sub-M$_{ch}$ explosion models for SNIa and they have shown that $X$-ray and 
$\gamma$-ray spectral evolution of both models are very different. 
In particular, they show that:
\begin{itemize}
\item The $\gamma$-ray light curves in sub-M$_{ch}$ models are much brighter 
and peak much earlier than in M$_{ch}$ SNIa. The $^{56}Ni$ $\gamma$-ray 
line emission (at 847 keV) from a bright sub-M$_{ch}$ explosion at 15 Mpc 
would be just at the limit for detection by the ESA satellite INTEGRAL 
(International Gamma Ray Astrophysics Laboratory) which will be launched in 
2001.
\item The presence of surface $^{56}Ni$ in sub-M$_{ch}$ SNIa would make them 
very bright emitters of iron peak K-shell emission visible for several 
hundred days after explosion. K-shell emission from a bright sub-M$_{ch}$ 
located near Virgo would be just above the limit for detection by the XMM 
observatory. 
\end{itemize}
\end{itemize}
Anyway, let us only notice that a first {\it true detection} 
of the 847 keV line of a true typical SNIa by INTEGRAL may help to better 
understand the SNIa models and the optical light curves of SNIa (see below 
2.3.3).
\subsubsection{On optical light curves of SNIa, {\it The Phillips Relation}}
For a long time, the optical light curve shapes of SNIa were generally supposed to be \underline{homogeneous} i.e. all SNIa were identical explosions with 
identical light curve shapes and peak luminosities (Woosley \& Weaver 
1986, see also our table 1). This was also supported by observations: for 
instance SN 1980N and SN 1981D in the galaxy NGC 1316 exhibited almost identical brightness and identical shapes of their light curves. It was this 
{\it uniformity} in the light curves of SNIa which had led to their primary 
use as {\it standard candles} in cosmology.
\\
Then, recent work on large samples of SNIa with high quality data has focused 
attention on examples of \underline{differences} within the SNIa class: in particular there are observed deviations in luminosity at maximum light, in colors, in light curve widths etc... However, despite all these differences, the 
discovery of some regularities in the light curve data has also emerged, in 
particular, the {\it Phillips relation}: {\it the brightest supernovae have 
the broadest light curve peaks}. In particular, see the upper plot of figure 2 in section 4. In effect, Phillips (1993) quantified the 
photometric differences among a set of nine well-observed SNIa using the 
following parameter $\Delta m_{15}(B)$ which measures the total drop, in 
$B$-magnitudes, from maximum to $t=15$ days after $B$ maximum. This 
{\it Phillips relation} is also called the {\it width luminosity 
relation (WLR)}. Finally, it is 
this emprical brightness decline relation (WLR) which allows the use of SNIa 
as {\it calibrated candles} in cosmology -see section 4- 
\\
Because SNIa are certainly more complex that can be described as a single-
parameter supernova family, several groups of theorists -experts in supernovae- 
have done a lot of work on SNIa in the early times and on their possible 
evolution until now -see most of the papers written in the proceedings edited 
by Niemeyer \& Truran (2000) and those edited by Holt \& Zhang (2000). But 
let us briefly recall how we can explain the present SNIa light curves and the 
{\it Phillips relation} or WLR. See, in particular, Woosley \& Weaver (in Petscheck, 1990), Arnett (1996), Nomoto et al. (1997) and more recently Pinto \& 
Eastman (2000a, 2000b, 2000c). 
\\
In theoretical models of light curves, the explosion energy goes into the kinetic energy of expansion E. The light curves are also powered by the radioactive decay chain:
$$
^{56}Ni \rightarrow ^{56}Co \rightarrow ^{56}Fe.
$$
The theoretical peaks of the light curve is at about 15-20 days after the explosion. Their decline is essentially due to the increasing transparency of the 
ejecta to $\gamma$-rays and to the decreasing input of radioactivity. The 
light curve shape depends mainly on the diffusion time: 
$$
\tau_D \sim \Bigl( \frac{\kappa M}{v_{esc} c} \Bigr)^{1/2}
$$
where $\kappa$ is the opacity, $v_{esc} \sim (E/M)^{1/2}$. Or, in other 
words, the theoretical light curve of a SNIa is essentially determined by 
three effects: 
\begin{itemize}
\item i) The deposition of energy from radioactive decays.
\item ii) The adiabatic conversion of internal energy to kinetic energy of 
expansion.
\item iii) The escape of internal energy as the observed light curve.
\end{itemize}
A priori. the light curve models of SNIa depends on, at least, four parameters: 1) the total mass $M$; 2) the explosion energy $E$; 3) the $^{56}Ni$ mass or 
$M_{56}$; 4) the opacity $\kappa$.
\\
Therefore, the question can be the following one : how these four (or five) parameters 
collapse to one for leading to the WLR ? Finally, one must only mention again 
the work of Pinto  \& Eastman (2000c) which shows that:
\begin{itemize}
\item 1) The WLR is a consequence of the radiation transport in SNIa.
\item 2) The main parameter is the mass of radioactive $^{56}Ni$ produced in 
the explosion and the rate of change of $\gamma$-ray 
escape.
\item 3) The small differences in initial conditions which might arise 
from evolutionary effects between $z\sim 0$ and $z \sim 1$ are 
unlikely to affect the SN cosmology results.
\end{itemize}
To conclude this section 2, one can say that although there are uncertainties 
in the nature and evolution of SNIa progenitors, in the ignition time and 
nature of the burning front, the empirical Phillips relation may be due to 
the radiation transport in SNIa and to atomic physics and not to hydrodynamics of the SNIa explosion (if the results of Pinto \& Eastman are confirmed).
\\
Anyway, in section 3, we will recall the cosmological background needed to 
use any standard candle for studying the geometry of the Universe -and in 
section 4, we will show how cosmologists use the Phillips relation to 
calibrate SNIa and use them as calibrated candles. 
\section{Cosmological Background}
In order to understand all the cosmological implications of observations of 
SNIa at high redshift, one must recall some theoretical background.
\subsection{The ($\Omega_m$,$\Omega_\Lambda$)-plane}
The cosmological term -i.e. the famous $\Lambda$ term- was introduced by 
Einstein when he applied General Relativity to cosmology:
\begin{equation}
G_{\mu \nu} = 8 \pi G T_{\mu \nu} + \Lambda g_{\mu \nu}
\end{equation}
where $g_{\mu \nu}$ is the metric, $G_{\mu \nu}$ is the Einstein tensor, 
$\Lambda$ is the cosmological constant and $T_{\mu \nu}$ is the energy-
momentum tensor such as $\nabla_\nu T^{\mu \nu}=0$.
\\ For a homogeneous and isotropic space-time, $g_{\mu \nu}$ is the Robertson-
Walker metric:
\begin{equation}
ds^2 = g_{\mu \nu} dx^{\mu} dx^{\nu} 
= dt^2 -R^2 \Bigl[\frac{dr^2}{1-kr^2} 
+r^2(d\theta^2 + sin^2 \theta \, d\phi^2 \Bigr]
\end{equation}
where $R(t)$ is the cosmic scale factor, $k$ is the curvature constant, 
$T_{\mu \nu}$ has the {\it perfect fluid} form: 
$$
T_{\mu \nu}= pg_{\mu \nu} + (\rho + p) u_{\mu} u_{\nu},
$$
with $\rho$, $p$, $u_\mu$ being respectively the energy density, the pressure 
and the 4-velocity:
$$
u_\mu = \frac{dx^\mu}{ds}.
$$
Therefore, the Einstein field equations (1) simplify in:
\begin{itemize}
\item The Friedman-Lemaitre equation for the expansion rate $H$, called the 
Hubble parameter:
\begin{equation}
H^2 \equiv \Bigl( \frac{\dot{R}}{R} \Bigr)^2
= \frac{8 \pi G \rho}{3} + 
\frac{\Lambda}{3} - \frac{k}{R^2}
\end{equation}
\item An equation for the acceleration:
\begin{equation}
\frac{\ddot{R}}{R} = - \frac{4 \pi G}{3} (\rho + 3p) + \frac{\Lambda}{3}
\end{equation}
where $\dot{R}=dR/dt$ and $k$ is the curvature constant which equals 
respectively to -1,0,+1 for a universe which is respectively open, flat and 
closed. Equation (3) says us that three competing terms drive the expansion: 
a term of energy (matter and radiation), a term of {\it dark energy} or 
cosmological constant, and a term of geometry or curvature term.
\end{itemize}
\underline{Remarks}: {\it i)} if one introduces the equation of state: 
$p=\omega \rho$, let us recall that $\omega_R=1/3$ for the radiation, 
$\omega_M=0$ for the cold matter, $\omega_\Lambda=-1$ for the cosmological 
constant. {\it ii)} Let us notice that one must, a priori, not only consider 
a static, uniform vacuum density (or cosmological constant) but also a 
dynamical form of evolving , inhomogeneous {\it dark energy} 
(or {\it quintessence}); 
for this last case: $-1<\omega_Q<0$.
\\
It is convenient to assign symbols to the various fractional constributions of 
the energy density at the present epoch such as:
\begin{equation}
\Omega_M = \frac{8 \pi G \rho_o}{3 H_o} , \ \ 
\Omega_\Lambda = \frac{\Lambda}{3 H_o} \ \ 
{\rm and} \ \ 
\Omega_k = \frac{k}{R^2_o H^2_o} 
\end{equation}
where the index $o$ refers to the present epoch. Then equation (3) can be 
written:
\begin{equation}
\Omega_m + \Omega_\Lambda + \Omega_k = 1
\end{equation}
and the astronomer's cosmological constant problem can be formulated by the 
following simple observational question:
\\
{\it Is a non-zero $\Omega_\Lambda$ required to achieve consistency in 
equation (6) ?}
\\
In comparison with $H_o$ and $\Omega_M$, the attempts 
to measure $\Omega_\Lambda$ are infrequent and modest in scope. 
Moreover signatures of $\Omega_\Lambda$ are more subtle 
than signatures of $H_o$ 
and $\Omega_M$, at least from an observational point of view. 
Anyway, studying possible variations in equation (6) having dominant versus 
negligible $\Omega_\Lambda$-term is quite challenging !
\\
Let us consider now the expansion dynamics in non-zero cosmological constant 
models. If:
\begin{equation}
a \equiv \frac{1}{1+z} = \frac{R}{R_o}
\end{equation}
is the expansion factor relative to the present epoch and if $\tau=H_o t$ is 
a dimensionless time variable (i.e. time in units of the measured Hubble 
time 1/$H_o$), the Friedmann equation (3) can be rewritten:
\begin{equation}
\Bigl( \frac{da}{d\tau} \Bigr)^2 = 1 + \Omega_M (\frac{1}{a} -1 ) + 
\Omega_\Lambda (a^2 -1)
\end{equation}
$\Omega_M$ and $\Omega_\Lambda$, here, are constant that parametrize the past 
(or future) evolution in terms of quantities of the present epoch.
\\
Equivalently, one can also parametrize the evolution with $\Omega_M$ and the 
deceleration parameter $q_o$:
\begin{equation}
q_o = - \Bigl( \frac{R\ddot{R}}{\dot{R}^2} \Bigr)_o
\end{equation}
which can also be written as:
\begin{equation}
q_o = \frac{\Omega_M}{2} - \Omega_\Lambda.
\end{equation}
Finally, the expansion dynamics can be given by the system:
\begin{equation}
\Bigl( \frac{\dot{a}}{a} \Bigr)^2 = \Omega_M (\frac{1}{a} )^3 + 
(1-\Omega_M - \Omega_\Lambda) (\frac{1}{a})^2 
+ \Omega_\Lambda
\end{equation}
\begin{equation}
\frac{\ddot{a}}{a} = \Omega_\Lambda - \frac{\Omega_M}{2} (\frac{1}{a})^3
\end{equation}
For different values of ($\Omega_M$, $\Omega_\Lambda$) one gets different 
expansion histories. Here, we only give the figure of the $(\Omega_M, 
\Omega_\Lambda)$-plane which displays various possible regimes -see figure 1 
of the ($\Omega_M,\Omega_\Lambda$)-plane. Of course, the results of high $z$-
supernova observations will be first represented by the Hubble diagram, but 
their analysis will be given in this ($\Omega_M,\Omega_\Lambda$)-plane where 
it can be confronted to other major recent cosmological observational results.

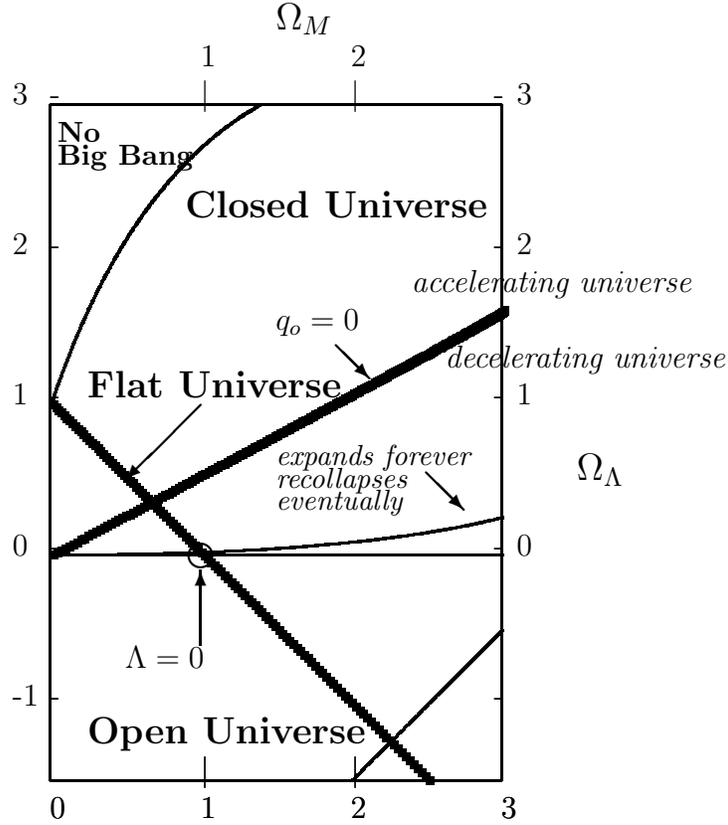
\begin{figure}[h]
\begin{center}
\setlength{\unitlength}{1.cm}
\begin{picture}(7,11)
  \thicklines

  \put(0,0){\line(1,0){6}}
  \put(0,0){\line(0,1){9}}
  \put(0,9){\line(1,0){6}}
  \put(6,0){\line(0,1){9}}
  \linethickness{3pt}
  \qbezier(0,3)(6,6.2)(6,6.25)
  \thicklines
  \qbezier(0,3)(4,3)(6,3.5)
  \qbezier(0,3)(3,3)(6,3)
  \qbezier(0,5)(1,8)(2.8,9)

  \put(2,3){\circle{0.3}}
  \put(2,1.8){\vector(0,1){1}}
  \put(1.8,7.5){{\bf \large Closed Universe}}
  \put(0.5,0.5){{\bf \large Open Universe}}
  \put(4.8,6.5){{\it accelerating universe}}
  \put(5.25,5.5){{\it decelerating universe}}
  \put(3,6){$q_o=0$}
  \put(3.8,5.8){\vector(1,-1){0.45}}
  \put(2,5){\vector(-1,-1){1}}
  \put(0.5,5.1){{\bf \large Flat Universe}}
  \put(3,4.2){\footnotesize \it expands forever}
  \put(3,3.9){\footnotesize \it recollapses}
  \put(3,3.6){\footnotesize \it eventually}
  \put(5,4.1){\vector(1,-1){0.5}}
  \put(1.0,1.5){$\Lambda=0$}
  \linethickness{3pt}
  \qbezier(0,5)(2,3)(5,0)
  \thicklines

  \put(3.,10.){{\large \bf $\Omega_M$}}
  \put(7,4.){{\large \bf $\Omega_\Lambda$}}
  \thicklines
  \put(0.1,8.5){\bf {\footnotesize No}}
  \put(0.1,8.2){\bf{\footnotesize Big Bang}}

  \qbezier(4,0)(6,2)(6,2)

  \put(2,9.5){1}
  \put(2,9){$|$}
  \put(4,9.5){2}
  \put(4,9){$|$}

  \put(0,-0.5){0}
  \put(2,-0.5){1}
  \put(2,0){$|$}
  \put(4,-0.5){2}
  \put(4,0){$|$}
  \put(6,-0.5){3}

  \put(0,-0.5){0}

  \put(2,-0.5){1}
  \put(2,0){$|$}
  \put(4,-0.5){2}
  \put(4,0){$|$}
  \put(6,-0.5){3}

  \put(0,1){-}
  \put(0,3){-}
  \put(0,5){-}
  \put(0,7){-}
  \put(0,9){-}
  \put(-0.5,1){-1}
  \put(-0.5,3){0}
  \put(-0.5,5){1}
  \put(-0.5,7){2}
  \put(-0.5,9){3}

  \put(5.9,1){-}
  \put(5.9,3){-}
  \put(5.9,5){-}
  \put(5.9,7){-}
  \put(5.9,9){-}
  \put(6.2,3){0}
  \put(6.2,5){1}
  \put(6.2,7){2}
  \put(6.2,9){3}
  
\end{picture}
\vskip2mm
\caption{: $(\Omega_M,\Omega_\Lambda)$-plane.}
\end{center}
\end{figure}

Of course, the most direct and theory independent way to measure $\Lambda$ would 
be to actually determine the value of the scale factor $a$ as a function of time.
 But, it is very difficult ! However with sufficiently precise information about 
the dependence of a distance-measure on $z$, one will try to disentangle 
the effects of matter, cosmological constant and spatial curvature.
\subsection{The luminosity distance $D_L$}
The Hubble diagram is a graphic representation of the luminosity  
distance, i.e. magnitude, of 
some class of objects -called standard candles- as a function of their redshift. 
Before presenting and discussing the recent results concerning type Ia-supernovae
 taken as {\it standard candles}, we recall some basic facts.
\subsubsection{On the luminosity distance $D_L$}
In cosmology, several different distance measurements are in use. They are all related by {\it simple} $z$-factors (see for instance Weinberg 1972, Peebles 1993, Hoggs 1999). The one which is relevant for our study is the luminosity-
distance:
\begin{equation}
D_L = \Bigl( \frac{{\mathcal L}}{4 \pi {\mathcal F}} \Bigr) ^{1/2}
\end{equation}
where ${\mathcal L}$ is the intrinsic luminosity of the source and ${\mathcal F}$, the 
observed flux. For Friedmann-Lemaitre models, one can show that:
\begin{equation}
D_L(z) = \frac{c}{H_o} d_L(z;\Omega_M ;\Omega_\Lambda) 
= D_H d_L
\end{equation}
where $D_H \equiv c/H_o$ is the Hubble distance and $d_L$   is a known 
dimensionless function of $z$, which depends parametrically on $\Omega_M$ and 
$\Omega_\Lambda$ defined by equations (5). 
In effect, the luminosity-distance $D_L$ is related to the transverse-comoving 
distance $D_m$ through:
\begin{equation}
D_L = (1 +z) D_m
\end{equation}
with:
\begin{equation}
\bullet \ D_m = D_H \Omega^{-1/2} \, {\rm sin} h [\Omega_k^{1/2} \, D_c/D_H], 
\ \ 
{\rm if} \ \ \Omega_k > 0
\end{equation}
\begin{equation}
\bullet \ D_m = D_c=D_H \, \int_0^z \, \frac{dz'}{E(z')}
\ \ {\rm if} \ \ \Omega_k = 0
\end{equation}
$$
{\rm with} \ \ 
E(z)= \Omega_m(1+z)^{3}+ 
\Omega_k(1+z)^2 + \Omega_\Lambda ]^{1/2}
$$
\begin{equation}
\bullet \ D_m = D_H \Omega^{-1/2} \, {\rm sin} [\Omega_k^{1/2} \, D_c/D_H],  
\ \ {\rm if} \ \ \Omega_k < 0.
\end{equation}
Therefore, from equations (6) and (14-18) we check that, for 
different possible values of $\Omega_k$:
\begin{equation}
D_L = D_L (z,\Omega_M, \Omega_\Lambda, H_o) 
\ \ {\rm and} \ \ 
d_L = d_L (z, \Omega_M, \Omega_\Lambda)
\end{equation}
\subsection{On the magnitude redshift relation}
The apparent magnitude $m$ of an object is related to its absolute 
magnitude $M$ through the distance modulus by:
\begin{equation}
m-M = 5 {\rm log}_{10} \, \frac{D_L}{{\rm Mpc}} + 25.
\end{equation}
Therefore, from equation (14-18) and (20) we obtain a relation 
between the apparent magnitude $m$ and $z$ with $\Omega_\Lambda$ and 
$\Omega_M$ as parameters:
\begin{equation}
m= 5 {\rm log}_{10}  \, d_L(z, \Omega_M, \Omega_\Lambda) + {\mathcal M}
\end{equation}
Let us only note here that:
\begin{itemize}
\item {\rm i)}${\mathcal M} \equiv M -5 {\rm log} \, H_o -25$ is a non-important 
fit parameter.
\item {\it ii)} It is the comparison of the theoretical expectation (21) with 
data which can lead to constraints on the parameters $\Omega_M$ and 
$\Omega_\Lambda$.
\end{itemize}
In other words and as we will see, the likelihoods for cosmological parameters 
$\Omega_M$ and $\Omega_\Lambda$ will be determined by minizing the $\chi^2$ 
statistic between the measured and predicted distances/magnitudes of SNIa 
taken as standard candles.
\section{Observations of SNIa at high redshift}
There are two major teams investigating high-$z$ SNIa:
\begin{itemize}
\item i) The {\it Supernova Cosmology Project} (SCP) which is led by 
S. Perlmutter (see S. Perlmutter et al. (1997, 1998, 1999 and references therein).
\item ii) The {\it High z-Supernovae Search Team} (HZT) which is led by B. Schmidt 
(see B. Schmidt et al. (1998 and all references therein). 
\\
Both groups have published almost identical results.
\end{itemize}
In Section (2), we briefly presented SNIa and emphasized that there are large 
uncertainties in the theoretical models of these explosive events as well as 
on the nature of their progenitors. We have also seen that although they 
cannot be considered as perfect {\it standard candles}, convincing evidence 
has been found for a correlation between light curve shape and luminosity of 
nearby SNIa -brighter implies broader- which has been quantified by Phillips 
(1993). In this section, we briefly present the cosmological use of SNIa at 
high redshift with: the observations, the results, a discussion of these 
results and their cosmological implications.  
\subsection{The observations}
Both groups -the SCP and the HZT- developed a strategy that garantee the 
discovery of many SNIa on a certain date. Here, we briefly review the 
strategy described by the SCP for its early campaigns.
\subsubsection{On the strategy}
Just after a new moon, they observed some 50-100 high galactic latitude 
fields -each containing about 1000 high-$z$ galaxies- in two nights, at 
the Cerro-Tololo 4m telescope in Chile with a Tyson and Bernstein's 
wide-field-camera. They returned three weeks later to observe the same 
fields. About two dozen SNIa were discovered in the image of the ten thousands 
of galaxies which were still brightening since the risetime of a SNIa is 
longer than 3 weeks. They observed the supernovae with spectroscopy at 
maximum light at the Keck-Telescope and with photometry during the following 
two months at the Cerro-Tololo International Observatory (CITO), Isaac 
Newton Telescope (INT) and -for the hightest $z$ SNIa- with the Hubble 
Space Telescope (HST).
\subsubsection{On spectra}
With the Keck (10m-telescope): {\it i)} they confirmed the nature of a large 
number of candidate supernovae; {\it ii)} they searched for peculiarities 
in the spectra that might indicate evolution of SNIa with redshift. In effect, 
as a supernova brightens and fades, its spectrum changes showing 
on each day which elements, in the expanding atmosphere, are passing through 
the photosphere. 
\\
In principle, their spectra must give a constraint on high-$z$ SNIa: they 
must show all of the same features on the same day after the explosion as 
nearby SNIa; or else, one has evidence that SNIa have evolved between the 
epoch $z$ and now.
\subsubsection{On light curves}

\begin{figure}[h]
\begin{center}
\epsfig{file=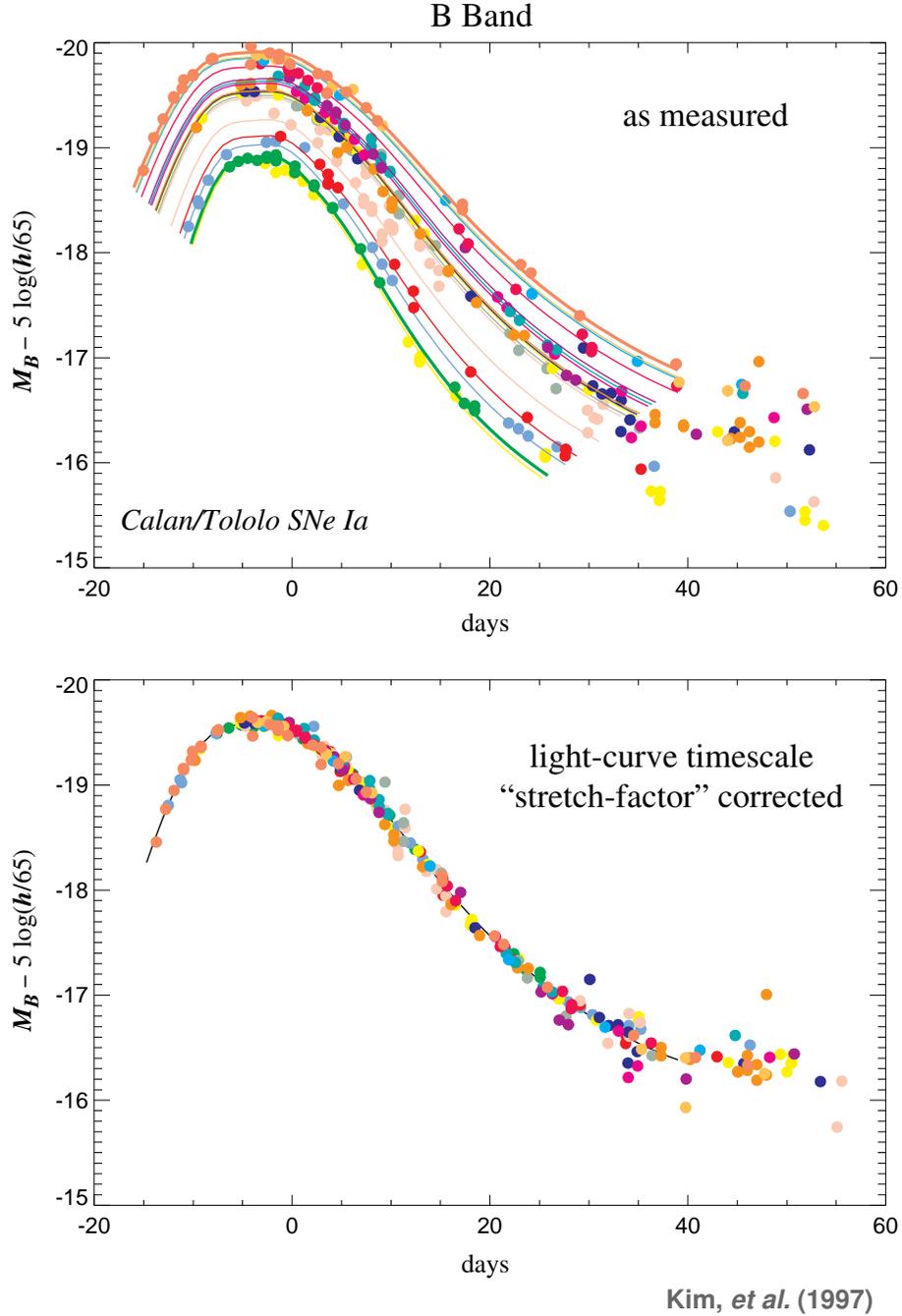,width=12cm,angle=0}
\caption{: The upper plot shows nearby SNIa light curves which 
exhibit LWR; the lower plot shows the nearby SNIa light curves, measured in 
the upper plot, after the {\it stretch parameter} correction.}
\label{fig:fig1}
\end{center}
\end{figure}

As already noted, in Section 2, nearby SNIa show a relationship between 
their peak luminosity and the timescale of their light curve: {\it 
brighter implies broader}. This correlation, often called {\it Phillips 
relation} or {\it luminosity width relation} (LWR) has been refined and 
described through several versions of a {\it one-parameter brightness decline 
relation}: {\it i)} $\Delta m(15)$ (Hamuy et al. 1996): {\it ii)} multicolor 
light curve shape (MLCS) (Riess et al. 1996); {\it iii)} Stretch parameter 
(Perlmutter et al. 1998). Moreover, there exists a simple linear relation 
between the absolute magnitude and the stretch parameter. Fig. (2) shows 
how the stretch correction aligns both the light curve width and the peak 
magnitude for the nearby Hamuy supernovae.
\subsubsection{On the analysis}
One can summarize the three-steps analysis of the 42 high-$z$ SNIa presented 
in Perlmutter et al. (1998). For each supernova:
\begin{itemize}
\item i) The final image of the host galaxy alone is substracted from the many 
images of the given SNIa spanning its light curve.
\item ii) Perlmutter et al. (1998) computed a peak-magnitude in the $B$-band corrected 
for galaxy extinction and the {\it stretch parameter} that stretches the time
 axis of a template SNIa (see 4.1.3) to match the observed light curve.
\item iii) Then, all of the SN magnitudes -corrected for the stretch-lumisosity relation- are plotted in the Hubble diagram as a function of their host 
galaxy redshift.
\end{itemize}
\subsection{The results and discussion}
One must compare the redshift dependence of observed $m$ with the theoretical
 expectation given above in Section 3:
\begin{equation}
m\, = \, 5 \, {\rm log}  \, d_L (z, \Omega_M, \Omega_\Lambda) + {\mathcal M}
\end{equation}
The effective magnitude versus redshift can be fitted to various cosmologies, 
in particular:
\begin{itemize}
\item {\it i)} to the flat universes ($\Omega_M + \Omega_\Lambda =1$)
\item {\it ii)} to the $\Lambda=0$-universes.
\end{itemize}
\subsubsection{On the Hubble diagram}

\begin{figure}[h]
\begin{center}
\epsfig{file=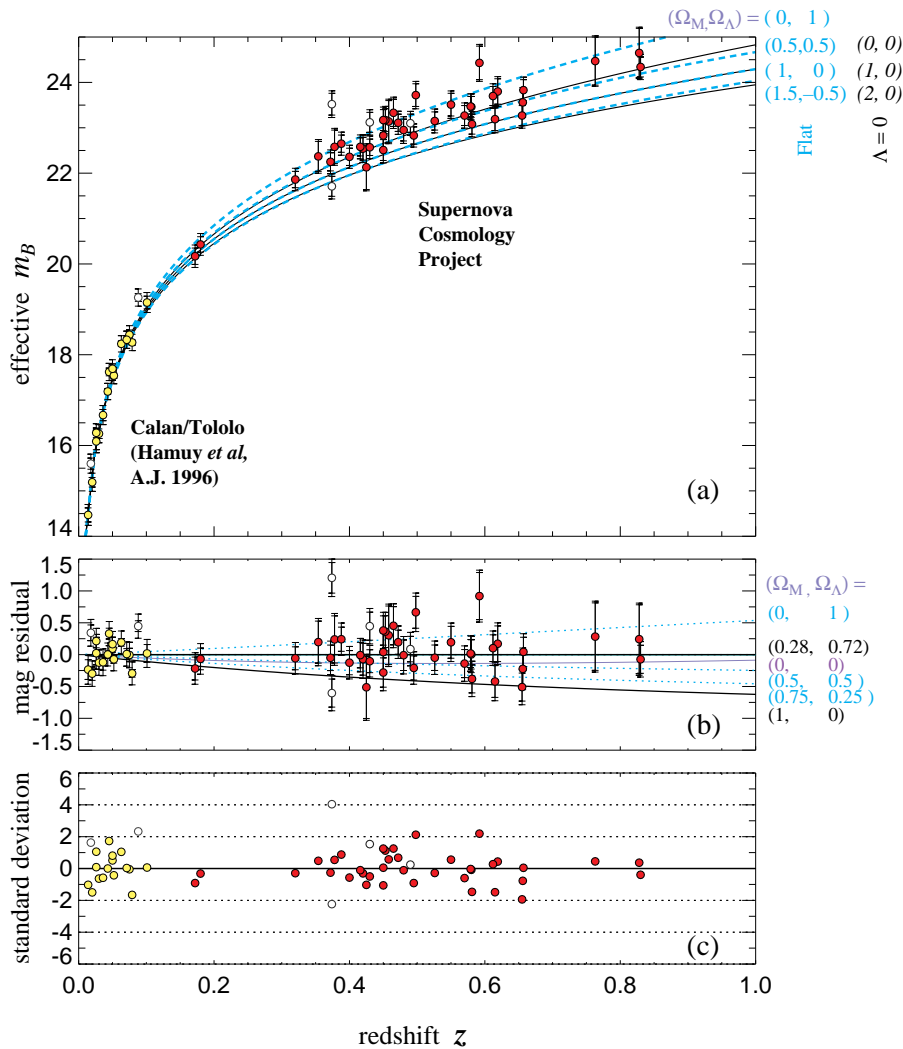,width=12cm,angle=0}
\caption{: From Supernova Cosmology Project (SCP): Hubble diagram.}
\label{fig:fig2}
\end{center}
\end{figure}

In Fig. (3), one sees the Hubble diagram for 42 high-$z$ SNIa from the SCP 
and 18 SNIa from the Cal\`an-Tololo Supernovae Survey -after correcting 
both sets for the LWR (see 4.1.4 above)- on a linear redshift scale 
to display details at high $z$. The solid curves are the theoretical 
$m^{eff}_B (z)$ for a range of cosmological models with $\Lambda=0$, 
$(\Omega_M,\Omega_\Lambda)=(0,0)$ on top: (1,0) in middle; (2,0) on bottom. 
The dashed curves are for flat cosmological models $(\Omega_M + 
\Omega_\Lambda =1)$: $(\Omega_M, \Omega_\Lambda)=(0,1)$ on top; 
(0.5,0.5) in middle; (1.5,0.5) on bottom. 
The best fit (Perlmutter et al. 1998) for a flat Universe $(\Omega_M + 
\Omega_\Lambda =1)$: $\Omega_M \sim 0.28$; $\Omega_\Lambda \sim 0.72$.
\\
The middle panel of Fig. (3) shows the magnitude residuals between data and 
models from the best fit flat cosmology: $(\Omega_M,\Omega_\Lambda) 
=(0.28, 0.72)$. The bottom panel shows the standard deviations. Note that, in their last 
Hubble diagram Kim (2000), Perlmutter (2000) extend it beyond $z=1$ by 
estimating the magnitude of SN1999eq ({\it Albinoni}). But, more analysis and more SNIa 
at $z>1$ are needed.  
\subsubsection{On the ($\Omega_M,\Omega_\Lambda)$-plane}

\begin{figure}[h]
\begin{center}
\epsfig{file=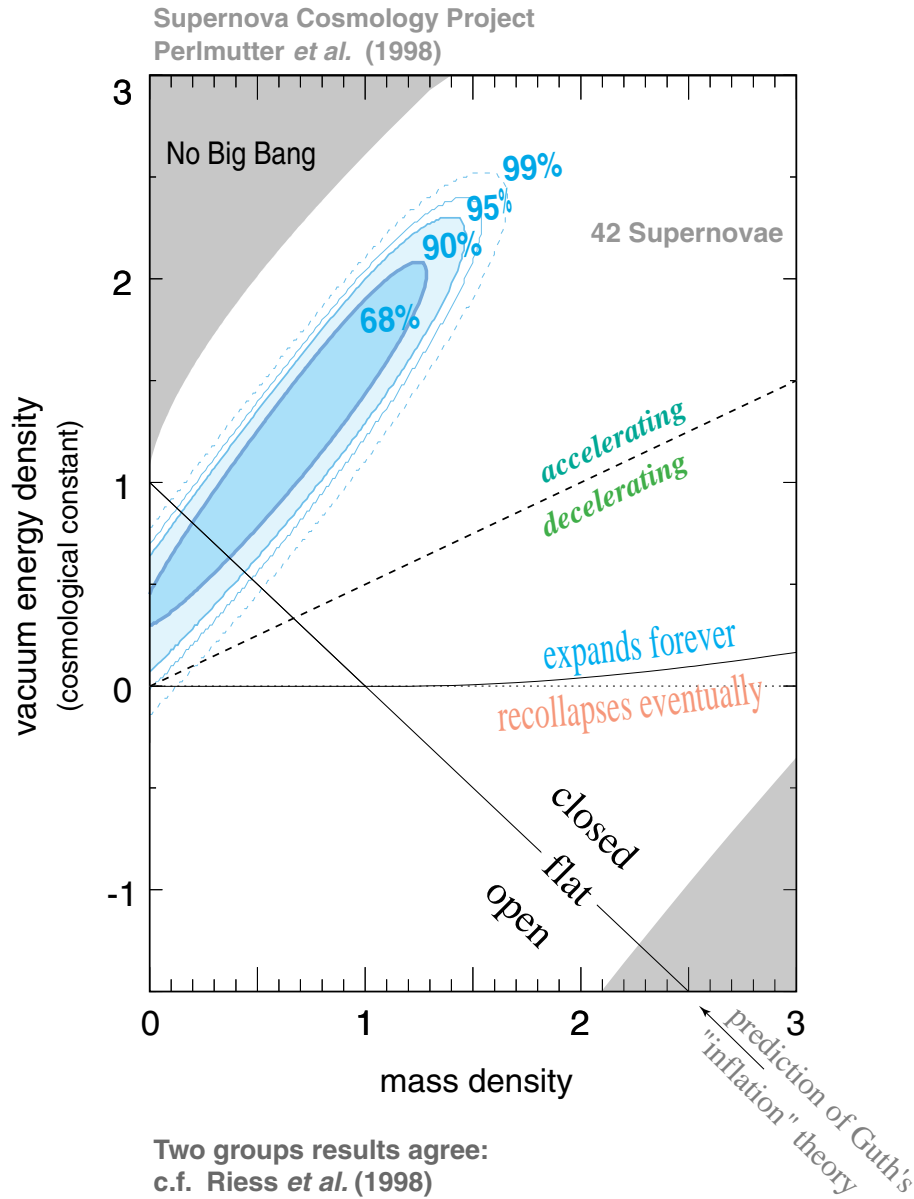,width=12cm,angle=0}
\caption{: From Supernova Cosmology Project (SCP): the ($\Omega_M, 
\Omega_\Lambda$)-plane.}
\label{fig:fig3}
\end{center}
\end{figure}

Fig. (4) shows the best fit confidence regions in the ($\Omega_M,
\Omega_\Lambda)$-plane for the preliminary analysis of the 42 SNIa fit 
presented in Perlmutter et al. (1998).
\\
Note that the spatial curvature of the universe -open, flat or closed- is 
not deterninative of the future of the universe's expansion, indicated by the 
near-horizontal solid line.
\\
From these figures, one can say that the data are strongly inconsistent with 
the $\Lambda=0$, flat universe model. If the simplest inflationary theories 
are correct and the universe is spatially flat, then the supernova data imply 
that there is a significant positive cosmological constant.
\\
The analysis of Perlmutter et al. (1998) suggested a universe with:
$$
0.8 \Omega_M -0.6 \Omega_\Lambda \, \sim \, -0.2 \pm 0.1
$$ 
If one assumes a flat Universe ($\Omega_M + \Omega_\Lambda =1$), the data 
imply:
$$
\Omega_M^{flat} \, = \, 0.28 ^{+0.09}_{-0.08} \ \ 
(1\sigma _{stat})^{+0.05}_{-0.04} \ \ ({\rm syst.})
$$
More recent data of both groups (SCP and HST) and other independent likelihood 
analysis of the SNIa data provide robust constraints on $\Omega_M$ and 
$\Omega_\Lambda$ consistent with those derived by Perlmutter et al. (1998). For 
a spatially flat universe, the supernova data require a non-zero 
cosmological constant at a high degree of significance ($\Omega_\Lambda \geq 
0.5$ at 95 \% confidence level).
\\
But before interpreting the results of the analysis of the SNIa data, one 
must try to discuss the believability of these results. 
\subsection{Discussion of the results}
The current SNIa data set already has statistical uncertainties that are 
only a factor of two larger than the identified systematic uncertainties. Here,
 let us only list the main indentified sources of systematic uncertainty and 
only mention some additional sources of systematic uncertainty 
(from Perlmutter 2000):
\begin{itemize}
\item {\it i)} Identified sources of systematic error and their estimated 
contribution to the uncertainty of the above measurements:
\begin{itemize}
\item Malmquist bias: $\delta M \sim 0.04$
\item $\kappa$-correction and cross filter calibration: $\delta M < 0.03$
\item Non-SNIa contamination: $\delta < 0.05$
\item Galaxy extinction: $\delta M < 0.04$
\item Gravitational lensing by clumped mass: $\delta M < 0.06$
\item Extinction by extragalactic dust:$\delta M \sim 0.03$
\end{itemize}
\item {\it ii)} Additional sources of systematic uncertainties:
\begin{itemize}
\item Extinction by {\it gray dust}
\item Uncorrected evolution: SNIa behaviour may depend on properties of its 
progenitor star or binary-star system. The distribution of these stellar 
properties may evolve in a given galaxy and over a set of galaxies.
\end{itemize}
\end{itemize}
Let us only notice that the control of statistical and systematic uncertainties is one of the main goals of the Supernovae Acceleration Probe (SNAP), a 
2-m satellite telescope proposed by Perlmutter (2000)\footnote{see 
{\texttt{http://snap.lbl.gov}}}. Such supernova searches, extended to greater $z$, can make a much precise determination of the luminosity distance as a 
function of $z$, $d_L(z)$. Note also that a very optimistic estimate of the 
limiting systematic uncertainty of one percent is expected for such supernovae 
searches. For a detailed discussion of the present state of all the systematic 
uncertainties we refer the reader to the paper of Riess (2000) and references 
therein.
\\
Anyway, the believability of the high-z SNIa results turn on the reliability 
of SNIa as one-parameter standard candles. 
The one-parameter is the rate decline of the light curve ({\it 
the brighter implies the broader}) -see subsection 2.3.3 above.
\\
For a while, due to the lack of a good theoretical understanding of the Phillips relation (what is the physical parameter ?), the supernova experts were 
not completely convinced by the high-$z$ SNIa observations and results. After 
several meetings and a lot of theoretical papers -see, in particular, 
Ruiz-Lapente et al. (1997)
, Niemayer \& Truran (2000), Holt \& Zhang (2000), and
 of course the papers of Pinto \& Eastman (2000a, 2000b, 2000c) -the results 
of (SCP) and (HZT) appear to supernova experts almost quite convincing...
Let us only recall the conclusion of Wheeler (summary talk of 
{\it Cosmic Explosions}, December 1999: {\it my personal answer to the question of 
whether the Universe is accelerating is \underline{probably yes}. My answer 
to the query of do we know for sure is \underline{not yet !}}.
\subsection{Implications of the results. The dark energy}
Therefore, if one accepts the high-$z$ SNIa results, the confidence levels 
in the $(\Omega_M-\Omega_\Lambda)$-plane 
-which are consistent for both groups (SCP and HZT)- 
favor a positive cosmological constant, an accelerating universe and strongly 
rule out $\Omega_\Lambda=0$ universes. In other words, the 
current results of high-$z$ SNIa observations suggest that the expansion of the Universe is accelerating indicating the existence of a cosmological constant 
or dark energy.
\\
There are a number of reviews on the various aspects of the cosmological 
constant. The classic discussion of the physics of the cosmological constant 
is by Weinberg (1989). More recent works are discussed, for instance, 
by Straumann (1999), Carroll (2000), Weinberg (2000) and references 
therein.
\\
In particular Weinberg (2000) formulates the two cosmological constant 
problems as:
\begin{itemize}
\item {\it i)} Why the vacuum energy density $\rho_v ({\rm 
or} \ \  \Omega_\Lambda)$ 
is not very large ?
\item {\it ii)} Why $\rho_v \ ({\rm or} \ \Omega_\Lambda)$ is not only 
small but also -as high-$z$ SNIa results seem indicate- of the same order 
of magnitude as the present mass density of the universe ?
\end{itemize}
The challenge for cosmology and for fundamental physics is to determine the 
nature of the dark energy. One possibility is that the dark energy consists 
of vacuum energy or cosmological constant. In this case -as already noted 
above in the remarks of the subsection 3.1- the equation of state of the 
universe is 
$$
\omega_\Lambda \, \equiv \, \frac{p}{\rho} = - 1.
$$
An other possibility is {\it quintessence} a time-evolving, spatially 
inhomogeneous energy component. For this case, the pressure is negative and 
the equation of state is a function of $z$, i.e.  $\omega_Q(z)$ is such that:
$$
-1 \, < \, \omega_Q(z) \, < \, 0.
$$
Let us only notice here that:
\begin{itemize}
\item {\it i)} these models of dark energy are based on nearly-massless scalar
 fields slowly rolling down a potential; most of them are called {\it tracker 
models} of quintessence for which the scalar field energy density can parallel  that of matter or radiation, at least for part of its history. As Steinhardt 
said (in the introductory talk of the Workshop on {\it String Cosmology}, 
Vancouver, August 2000): {\it There are dumb trackers} 
(Steinhardt et al. 1998), {\it smart trackers} (Armandariz-Picon et al. 
2000a, 2000b; Bean \& Magueijo 2000). 
\item {\it ii)} there are other models of dark energy besides those based 
on nearly-massless scalar fields; for instance, models based on networks 
of tangled cosmic strings for which the equation of state is: 
$\omega_{string}=-1/3$ or based on walls for which $\omega_{wall}=-2/3$ 
(Vilenkin 1984, Spergel \& Pen 1996, Battye et al. 1999).
\end{itemize}
Anyway, a precise measurement of $\omega$ today and its time variation could 
provide important information about the dynamical properties and the nature 
of dark energy. Telescopes may play a larger role than accelerators in 
determining the nature of the dark energy. In this context, by observing 
SNIa at $z$ about 1, SCP and HZT have found strong evidence for dark energy 
i.e. a smooth energy component with negative pressure and negative equation 
of state. For the future -and as already noted in the subsection 4.2.3- the 
spatial mission SNAP is being planned; it is expected that about 1000-2000 
SNIa at greater $z$ ($0.1<z<1.7$) will be detected.
\\
SNAP has also very ambitious goals: it is expected that SNAP will test the 
nature of the dark energy, will better determine the equation of state of the 
universe and its time variation i.e. $\omega(z)$. But very recent studies (for 
instance Maor et al. 2000) show that the determination of the equation of 
state of the Universe by high-$z$ SNIa searches -by using luminosity 
distance- are strongly limited.   
\section{Conclusions}
After a brief review of type Ia supernovae properties (Section 2) and of 
cosmological background (section 3), we presented the work done by two 
separate research teams on observations of high-$z$ SNIa (Section 4) and 
their interpretation (Section 5) that the expansion of the Universe is 
accelerating. If verified, this will proved to be a remarkable discovery. 
However, many questions without definitive answers remain. 
\\
What are SNIa and SNIa progenitors ? 
\\
Are they single or double degenerate stars ? 
\\
Do they fit Chandrasekhar or sub-Chandrasekhar mass models at explosion ? 
\\
Is the mass of radioactive $^{56}Ni$ produced in the explosion really 
the main parameter underlying the {\it Phillips relation} ? 
\\
If so, how important are SNIa evolutionary corrections ? 
\\
Can systematic errors be negligible and can they be converted into 
statistical errors ? 
\\
Can a satellite, like SNAP, provide large homogeneous 
samples of very distant SNIa and help determine several key cosmological 
parameters with an accuracy exceeding that of planned CMB observations ?
\vskip2mm
\noindent
Meanwhile, cosmologists are inclined to believe the SNIa results, because of the 
preexisting evidence for {\it dark energy} that led to the prediction of 
accelerated expansion.   
\vskip2mm
\noindent
{\bf Acknowledgements}\\
The authors gratefully acknowledge Stan Woosley for helpful suggestions. 
Part of the work of D.P. has been conducted 
under the auspices of the {\it Dr Tomalla Foundation} and Swiss National 
Science Foundation.
\section*{References}
Armandariz-Picon C. et al. 2000a, {\texttt{astro-ph/0004134}
\\ 
Armandariz-Picon C. et al. 2000b, {\texttt{astro-ph/0006373}
\\
Arnett W.D. 1996 in {\it Nucleosynthesis and supernovae}, Cambridge Univ. Press
\\
Battye R. et al. 1999, {\texttt{astro-ph/9908047}
\\
Bean R., Magueijo J. 2000, {\texttt{astro-ph/0007199}
\\
Branch D. 1999 ARAA 36, 17
\\
Carroll S. 2000, {\texttt{astro-ph/0004075}
\\
Hamuy M. et al. 1996, AJ 112, 2398
\\
Fillipenko A., Riess A. 1999, {\texttt{astro-ph/9905049}
\\
Fillipenko A., Riess A. 2000, {\texttt{astro-ph/0008057}
\\  
Hoggs D. 1999, {\texttt{astro-ph/9905116}
\\
Holt S., Zhang W. 2000 in {\it Cosmic Explosions}, AIP in press
\\
Kim A. et al. 1997, {\texttt{http://www-supernova.lbl.gov}
\\
Kim A. et al. 2000, {\texttt{http://snap.lbl.gov}
\\
Livio M. et al. 2000, {\texttt{astro-ph/0005344}
\\
Maor I. et al. 2000, {\texttt{astro-ph/0007297}
\\
Niemeyer J., Truran J. 2000, in {\it Type Ia Supernovae: Theory and Cosmology} 
Cambridge Univ. Press
\\
Nomoto K. et al. 1984, ApJ 286, 644
\\
Nomoto K. et al. 1997, {\texttt{astro-/ph/9706007}
\\
Nugent et al. 1997 ApJ 485, 812
\\
Peebles J. 1993 in {\it Principles of Physical Cosmology}, Princeton 
Univ. Press
\\
Perlmutter S. et al 1997 ApJ 483, 565
\\
Perlmutter S. et al. 1998 Nature 391, 51
\\
Perlmutter S. et al. 1999 ApJ 517, 365
\\
Perlmutter S. 2000, {\texttt{http://snap.lbl.gov}
\\
Petschek A. 1990 in {\it Supernovae}, Springer-Verlag
\\
Phillips M. 1993 BAAS 182, 2907
\\
Pinto P., Eastman 2000a ApJ 530, 744
\\
Pinto P., Eastman 2000b ApJ 530, 756
\\
Pinto P., Eastman 2000c, {\texttt{astro-ph/0006171}
\\
Pinto P. et al. 2000, {\texttt{astro-ph/0008330} 
\\
Riess A. et al. 1996, ApJ 473, 588
\\
Riess A. 2000, {\texttt{astro-ph/0005229}
\\
Ruiz-Lapente R. et al. 1997 in {\it Thermonuclear supernovae}, Kluwer eds
\\
Schmidt B. et al. 1998 ApJ. 507, 46
\\
SNAP (Supernova Acceleration Probe), {\texttt{http://snap.lbl.gov}
\\
Spergel D., Pen U. 1997 ApJ 491, L67
\\
Steinhardt P. 1998, {\texttt{astro-ph/9812313}
\\
Straumann N. 1999 Eur. J. Phys. 20, 419 
\\
Vilenkin A. 1984, Phys. Rev. Lett. 53, 1016
\\ 
Weinberg S. 1972 in {\it Gravitation and Cosmology}, Wiley Eds
\\
Weinberg S. 1989, Rev. Mod. Phys. 61, 1
\\
Weinberg S. 2000, {\texttt{astro-ph/0005265}
\\
Wheeler S. 1999, {\texttt{astro-ph/9912403}
\\
Woosley S., Weaver T. 1986 ARAA 24, 205
\\
Woosley S., Weaver T. 1994 in {\it Les Houches, session LIV}, S. Bludman et al. Elsevier
\end{document}